\documentstyle[prd,aps]{revtex}

\begin{document}

\draft
\input epsf
\twocolumn[\hsize\textwidth\columnwidth\hsize\csname
@twocolumnfalse\endcsname

\title{Constraining the variation of the coupling constants \\
with big bang nucleosynthesis}
\author{Kazuhide Ichikawa and M. Kawasaki}
\address{Research Center for the Early Universe, University 
  of Tokyo, Bunkyo-ku, Tokyo 113-0033, Japan}

\maketitle

\begin{abstract}
We consider the possibility of the coupling constants of the $SU(3)\times SU(2)\times U(1)$ gauge interactions at the time of big bang nucleosynthesis having taken different values from what we measure at present, and investigate the allowed difference requiring the shift in the coupling constants not violate the successful calculation of the primordial abundances of the light elements. We vary gauge couplings and Yukawa couplings (fermion masses) using a model in which their relative variations are governed by a single scalar field, dilaton, as found in string theory. The results include a limit on the fine structure constant $-6.0\times10^{-4}<\Delta\alpha_{EM}/\alpha_{EM}<1.5\times10^{-4}$, which is two orders stricter than the limit obtained by considering the variation of $\alpha_{EM}$ alone. 
\end{abstract}

\vskip2pc]

\section{Introduction}
Big bang nucleosynthesis (BBN) is one of the most important tools for probing the early universe. The standard big bang nucleosynthesis (SBBN) predicts the primordial abundances of light elements (D, $^3$He, $^4$He and $^7$Li) with the only free parameter, the baryon-to-photon-ratio $\eta$. For $\eta\sim O(10^{-10})$, they are remarkably consistent with the observations of the abundance which extend over 9 digits. Requiring not to vitiate this consistency, the stringent constraints on various theories which affect cosmology have been obtained.

Among such theories constrained by BBN, there are some in which values of coupling constants may  vary. The SBBN prediction assumes, along with three light neutrino species and no lepton asymmetry, that physical parameters involved in the calculation (the fine structure constant $\alpha_{EM}$, the Fermi constant $G_{F}$, the electron mass $m_e$, etc.) are the same for now and the BBN time. However, this is not an obvious choice because there are numbers of ways in which they could have varied. In addition to the theoretical possibility, the recent analysis of quasar absorption lines found possible evidence for variation in $\alpha_{EM}$ \cite{QSO}.

In this paper, we consider a model taken from string theory where these coupling constants are all related to the expectation values of dilaton field $\Phi$ and could in principle vary with time. For example, very general arguments indicate that the dilaton cannot be stabilized at a value we would characterise as corresponding to weak coupling \cite{DineSeiberg85}. Also, in certain popular models for stabilising the dilaton using gaugino condensates, the cosmological evolution would almost inevitably tend to overshoot the desired minimum of the dilaton potential and run off to an anti-de Sitter vacuum \cite{BrusteinSteinhardt93}.    

The notion that the coupling constants are determined by a single scalar field not only motivates the investigation of their time variation as a probe for physics beyond the Standard Model, but also makes the analysis of simultaneous changes in the couplings simple and concrete. In addition to the free parameter of the SBBN $\eta$, we have only another parameter $\displaystyle\frac{\Delta\Phi}{\Phi}\equiv\frac{\Phi_{BBN}-\Phi_{now}}{\Phi_{now}}$, the fractional variation of the dilaton field. After we calculate the primordial abundances with various values of these two quantities, we search a parameter region not excluded by the observation. Thus obtained limit on $\Delta\Phi/\Phi$ is readily translated in the limits on the coupling constants.

Our analysis is positioned as the extension to Ref.~\cite{CampbellOlive}. There, the dilaton dependence of the coupling constants is determined by the action of the heterotic string in the Einstein frame. The constraint is obtained from $^4$He whose abundance and coupling dependence can be estimated without recourse to a numerical calculation. Their method is easy to calculate and appropriate for an order estimation. But we can extract more information by performing the numerical calculation, whose advantages are: 1.~the abundances of the light elements other than $^4$He are calculated so we can use more observational data, especially D's, to put a constraint, 2.~we can take into account of $\eta$ dependence of the abundances, 3.~we use realistic values for the reaction rates so we avoid the rough estimation of the weak reaction rate (both absolute value and temperature dependence) based on dimensional analysis, 4.~by Monte-Carlo simulation, we can estimate theoretical uncertainty, 5.~by calculating a statistical measure (such as $\chi^2$), we can objectively quantify the constraint. Of course, as the SBBN becomes more and more precise owing to the progress in the studies of nuclear reaction rates and of primordial abundances, numerical computation is necessary in general to reflect the recent development and to produce a result close to the truth. 

In the next section, we review briefly the SBBN and see what kind of physical quantities are needed to make the prediction. In Sec.~\ref{sec:coupling}, we estimate their coupling dependences. In Sec.~\ref{sec:dilaton}, we introduce a model containing the dilaton and we investigate the dilaton dependence of the coupling constants. In Sec.~\ref{sec:constraint}, we put a bound on $\Delta\Phi/\Phi$ by calculating $\chi^2$ with observational and theoretical uncertainty from the recent data (the former very much dominates). In terms of the limit on $\Delta\alpha_{EM}/\alpha_{EM}$, we find
\begin{equation}
-6.0\times10^{-4}<\frac{\Delta\alpha_{EM}}{\alpha_{EM}}<1.5\times10^{-4},
\end{equation} 
two orders more restrictive than the one found in Ref.~\cite{BIR} where only the variation of $\alpha_{EM}$ is taken into account. We discuss this originates from the unified gauge couplings, which relates electromagnetic sector to strong sector. At the energy scale of BBN, manifest quantities are $\alpha_{EM}$ and $\Lambda_{QCD}$ which appears instead of $\alpha_{strong}$ through dimensional transmutation. The former linearly depend on the dilaton but the latter does exponentially. $^4$He abundance is determined by the magnitude of neutron-proton mass difference $\Delta m$ which is in turn determined by $\Lambda_{QCD}$ and this dominates the constraints by BBN. 

\section{Physical quantities in the SBBN}\label{sec:SBBN}

In order to see how the effects of changing coupling constants arise, we summarise the main points of the SBBN calculation. Following Ref.~\cite{EarlyUniverse}, we divide it into three stages. The actual calculation is performed by computer which runs the code to solve the set of ordinary differential equations so, of course, does not distinguish such stages but we see the important quantities which determine the primordial abundances of the light elements appear in this brief review.

{\it 1st Stage (statistical equilibrium)}($T\gg1$ MeV; $t\ll1$ sec)

The energy density and the number density are dominated by the relativistic electron $e^-$, positron $e^+$, neutrino $\nu$, anti-neutrino $\bar{\nu}$ and photon $\gamma$. There are three types  of neutrino and anti-neutrino. There is only tiny fraction of protons and neutrons, just $\sim 10^{-9}$ number fraction. In this period, all these particles make elastic scattering so frequently that they are in thermal equilibrium and has equal temperatures. Especially, weak interaction cross sections are large and neutrinos are in thermal equilibrium with the other particles. In addition, there are weak interaction processes interchanging protons and neutrons,
\begin{eqnarray}
n+\nu_e \leftrightarrow p+e^-  \label{eq:ntop1} \\
n+e^+ \leftrightarrow  p + \bar{\nu}_e \label{eq:ntop2}\\
n \leftrightarrow p + e^- + \bar{\nu}_e \label{eq:ntop3},
\end{eqnarray}
so they are also in the chemical equilibrium. Then, the number ratio of proton to neutron at temperature $T$ is
 $\frac{n_n}{n_p}=e^{-(m_n-m_p)/T}e^{(\mu_n-\mu_p)/T}$ where $m_n (m_p)$ is neutron (proton) mass and $\mu_n(\mu_p)$ is its chemical potential, but the last factor $e^{(\mu_n-\mu_p)/T}$ is almost unity if the universe has no lepton asymmetry, which
is the assumption of SBBN. So as long as equilibrium holds with temperature $T$, the neutron-to-proton ratio is determined by the neutron-proton mass difference $\Delta m=m_n-m_p$,
\begin{equation}
\frac{n_n}{n_p} = e^{-\Delta m/T}. \label{eq:np-equiv-ratio}
\end{equation}

At this stage, the nuclear reaction rates are also fast enough for light nuclei to be in chemical equilibrium. Then the abundances of the nuclear species with mass number $A$ and atomic number $Z$ is
\begin{eqnarray}
Y_A&=&g_A [\zeta (3)^{A-1}\pi^{(1-A)/2}2^{(3A-5)/2}] A^{5/2}\left(\frac{T}{m_N}\right)^{3(A-1)/2} \nonumber \\
& &\times\eta^{A-1}Y_p^Z Y_n^{A-Z}\exp(B_A/T), \label{eq:NSE}
\end{eqnarray}
where $\zeta(3)\sim 1.202$, $m_N$ is the nucleon mass, $B_A$ is the binding energy and $g_A$ is the number of degrees of freedom. Since $\eta$ is small ($\sim 10^{-10}$) and $B_A\ll T$, nuclei abundance is negligible at this temperature.

{\it 2nd  Stage (neutron-proton freeze-out)} ($T\sim 0.7$ MeV; $t\sim 2$ sec)

As the universe expands and cools, the rates of interactions involving neutrinos decreases. Especially, there are some point when the reactions (\ref{eq:ntop1})-(\ref{eq:ntop3}) practically stop and the chemical equilibrium break down. After this point, the numbers of neutrons and protons do not change (actually, small amount of neutrons turns into protons by the beta decay) and their ratio is fixed (froze out) at the value determined from Eq.(\ref{eq:np-equiv-ratio}) with the temperature of that epoch. We call this "freeze-out temperature" and denote $T_f$.

The freeze-out occurs when the rate of the reaction $\Gamma(T)$ and the expansion rate of the universe $H(T)$ becomes equal:
\begin{equation}
H(T_f) \approx \Gamma_{weak}(T_f). \label{eq:focondition}
\end{equation}
The expansion rate is determined by the energy density $\rho$ through the Friedman equation,
\begin{equation}
H^2=\frac{8\pi G\rho}{3}. \label{eq:hubble}
\end{equation}
At this stage, $\rho$ is dominated by photons, neutrinos and relativistic electrons ($T>m_e=0.511$ MeV) and hence the cosmic density is given by $\rho=10.75(\pi^2/30)T^4$. The rate of weak interaction is 
\begin{eqnarray*}
\Gamma_{weak}&=&\Gamma(n\rightarrow p) \\
&=&\Gamma(n+\nu_e \rightarrow p+e^-) \\
& &{}+\Gamma(n+e^+ \rightarrow  p + \bar{\nu}_e)+\Gamma(n \rightarrow p + e^- + \bar{\nu}_e) .
\end{eqnarray*}
Using the Fermi theory of the weak interaction, this can be written as
\begin{eqnarray*}
\lefteqn{\Gamma(n\rightarrow p)}\hspace{0.3cm} \nonumber \\
&=&A\left[\int_0^{\infty}+\int_{-\infty}^{-\Delta m-m_e}+\int_{-\Delta m+m_e}^{0} \right]  dx\nonumber\\
& &\times\left(1-\frac{m_e^2}{(x+\Delta m)^2}\right)^{\frac{1}{2}} (x+\Delta m)^2 x^2\nonumber\\
& &\times \frac{1}{1+e^{x/T_{\nu}} } \frac{1}{1+e^{-(x+\Delta m)/T_e}} .\label{eq:totalweakrate}
\end{eqnarray*}
We calculate the normalization factor $A$ by using $\Gamma(n \rightarrow p + e^- + \bar{\nu}_e) |_{T=0}=\tau_n^{-1}$ where $\tau_n$ is the neutron lifetime. Integration gives $A=0.05606\ \tau_n^{-1}$ MeV$^{-4}$. From these expression and Eq.(\ref{eq:focondition}), we find $T_f\approx 0.7$ MeV and, from Eq.~(\ref{eq:np-equiv-ratio}), freeze-out ratio $\left({n_n}/{n_p}\right)_f = e^{-\Delta m/T_f}\approx 0.158$.

For the nuclear reaction, the situation did not change from the previous stage. Their abundances are still very small.

{\it 3rd Stage (light-element synthesis)} (0.7 MeV $> T > $ 0.05 MeV; 3 sec $<t<$ 6 min) 

After the second stage, the electron-positron pair annihilation completed so there remains the photons and the neutrinos as relativistic particles. 

Since the early universe has lower density than inside of stars, there practically occurs only two body reactions. Thus unless the deuterons are synthesized by the reaction $p+n\rightarrow {\rm D}+\gamma$, larger nuclei are not synthesized. This reaction does not proceed until the photon density is low enough not to photodissociate D or, in other words, the factor $\eta\exp(B_{\rm D}/T)$ in Eq.(\ref{eq:NSE}) becomes $O(1)$. It occurs at about $T=0.06$ MeV. After that, mainly charged-particle reactions such as ${\rm D}+{\rm D}\rightarrow t+p$ and $t+{\rm D} \rightarrow{\rm ^4He}+n$ proceed synthesizing almost all of the neutrons into $^4$He because it has the largest binding energy among the light elements. Therefore, $^4$He abundance (conventionally expressed by mass ratio) can be estimated by the frozen n-p ratio found at the 2nd stage:
\begin{equation}
Y_{^4{\rm He}}\approx\frac{2}{1+(n_p/n_n)_f}=\frac{2}{1+e^{\Delta m/T_f}} \label{eq:He4abundance}
\end{equation}
Besides $^4$He, small amount of D and $^3$He and very small amount of $^7$Li are synthesized. Their abundances are determined by $\eta$ and the reaction rates.

In summary, to perform BBN calculation, we need to input following values {\it at BBN time}:
 \begin{itemize}
\item neutron-proton mass difference $\Delta m$
\item neutron lifetime $\tau_n$
\item nuclear reaction rates
\end{itemize}
Mainly, the abundance of $^4$He are affected by the first two as seen from Eqs.~(\ref{eq:focondition}) and (\ref{eq:He4abundance}), and D, $^3$He and $^7$Li by the last one.

\section{The coupling constants dependence of the BBN inputs} \label{sec:coupling}
We know the present value of the BBN input quantities pointed out in the previous section with some experimental uncertainty. To estimate their values during the BBN, we try to express these quantities in terms of the coupling constants.  

\subsection{neutron-proton mass difference $\Delta m$} \label{sec:np-massdif}
The origin of the neutron-proton mass difference is traced to the electromagnetic self-energy difference and the d-u quark mass difference (QCD chiral symmetry breaking by the mass terms). The former makes proton heavier than neutron but the latter does the inverse and the total is measured accurately to be $\Delta m\equiv m_n-m_p=1.2933318\pm0.0000005$ MeV \cite{PDG2000}. 

As argued in Ref.~\cite{quarkmasses} the largest contribution to the electromagnetic part comes from the Born term of the Cottingham formula \cite{Cottingham} and hence it can be calculated with  relatively less uncertainty. This formula expresses the nucleon self-energy in the first order of $\alpha_{EM}$ by its electric and magnetic form factor. Using the formula and experimental data for the form factors, the difference in the neutron and the proton is $-0.76$MeV. This value is proportional to (of course)  $\alpha_{EM}$ and to $\Lambda_{QCD}$ to have a proper dimension.

On the other hand, the absolute values of u and d quark masses are not well known,   $m_u=1\sim 5$ MeV and $m_d=3\sim 9$ MeV \cite{PDG2000}. But knowing the electromagnetic contribution, the quark contribution should be 2.05 MeV. This is proportional to the Yukawa couplings $y_u, y_d$ and the Higgs expectation value $\langle H\rangle$.

Now we can write 
\begin{eqnarray}
\Delta m&=&a\alpha_{EM}\Lambda_{QCD}+b(y_d-y_u)\langle H\rangle,
\end{eqnarray}
where $a$ and $b$ are constants we assume not to depend on any coupling constants. Thus, we estimate the mass difference at BBN epoch $\Delta m_{bbn}$ as
\begin{eqnarray}
\lefteqn{\Delta m_{bbn}}\hspace{0.3cm}\nonumber \\
&=&a\alpha_{EMbbn}\Lambda_{QCDbbn}+by_{bbn}\langle H\rangle_{bbn} \nonumber \\
&=&-0.76\frac{\alpha_{EMbbn}}{\alpha_{EMnow}}\frac{\Lambda_{QCDbbn}}{\Lambda_{QCDnow}}+2.05\frac{y_{bbn}}{y_{now}}\frac{\langle H\rangle_{bbn}}{\langle H\rangle_{now}} \ [{\rm MeV}].\nonumber \\
& & 
\end{eqnarray}

\subsection{neutron lifetime}
Neutron $\beta$-decay $n\rightarrow p+e+\bar{\nu}$ is very well approximated by the one-point interaction of four particles: neutron, proton, electron and neutrino. Its coupling constant is denoted $G_F$ called the Fermi coupling constant. Using this theory,
\begin{eqnarray}
\tau_n^{-1}&\propto& G_F^2\int d^3p_e d^3p_{\nu}\ \delta(E_e+E_{\nu}+m_p-m_n) \nonumber \\
&=&G_F^2\ 16\pi^2\int_{m_e}^{\Delta m} dE_e\ E_e\sqrt{E_e^2-m_e^2}\ (\Delta m-E_e)^2 \nonumber \\
&=&G_F^2\ 16\pi^2 m_e^5\ \frac{1}{60}\Bigl\{\sqrt{q^2-1}(2q^4-9q^2-8)\nonumber \\
& &{} +15q\log(q+\sqrt{q^2-1})\Bigl\}
\end{eqnarray}
where we defined $q\equiv\frac{\Delta m}{m_e}$. Denoting the factor $\frac{1}{60}\{\dots\}\equiv f(q)$,
\begin{eqnarray}
\tau_n &\propto& G_F^{-2} m_e^{-5} f(q)^{-1}\nonumber \\
&=&\langle H\rangle^4(y_e \langle H\rangle)^{-5} f(q)^{-1} =\langle H\rangle^{-1}y_e^{-5}f(q)^{-1}.
\end{eqnarray}
The first equality follows from $G_F=\frac{g_2^2}{M_W^2}=\frac{g_2^2}{(g_2\langle H\rangle)^2}=\frac{1}{\langle H\rangle^2}$ where $g_2$ is the SU(2) coupling constant and $M_W$ is the weak boson mass. Therefore, we obtain
\begin{eqnarray}
\tau_{n,bbn}=\left[\frac{y_{e,bbn}}{y_{e,now}}\right]^{-5}\left[\frac{\langle H\rangle_{bbn}}{\langle H\rangle_{now}}\right]^{-1}\left[\frac{f(q_{bbn})}{f(q_{now})}\right]^{-1} ,
\end{eqnarray}
where $y_e$ is the electron Yukawa coupling and $f(q_{now})=1.63615$.

\subsection{charged-particle induced reaction rates} \label{sec:charged}

Most of the reaction rates involved in the BBN calculation are charged-particle induced reaction rates. Their $\alpha_{EM}$ dependence is considered in Ref.~\cite{BIR} for the rates discussed in Ref.~\cite{SKM}. We implemented the $\alpha_{EM}$ dependence in the same manner updating the reaction rates recently compiled by Angulo et al.~\cite{NACRE}. Since only a few reaction rates have resonance terms which is considered to be dependent of the strong coupling constant, or in other words since the charged-particle reaction rates influential for BBN are practically determined by Coulomb barrier penetrability, we expect that neglecting their strong coupling dependence will not affect the results a lot.

\subsection{neutron-induced reaction rates}
The cross section for $n+p\rightarrow {\rm D}+\gamma$ is calculated at energies relevant to BBN using the effective field theory that describes the two-nucleon sector \cite{ChenSavage99}. The rate  obtained by thermal averaging this theoretical cross section reproduces the abundance calculation using the rate shown in Ref.~\cite{SKM}. We exploit this theoretical formula to estimate the coupling dependence by assuming the parameters which have the dimension [length]$^n$ in the formula proportional to $m_{\pi}^{-n}$. The coupling dependence of pion mass $m_{\pi}$ is known from Gell-Mann-Oakes-Renner relation to be $m_{\pi}^2\propto m_{quark}\Lambda_{QCD}$. The other parameters in the formula are the nucleon mass $m_N$ and the deuteron binding energy $B_D$. Their dependence is estimated by $m_N\propto\Lambda_{QCD}$ and $B_D\propto\frac{m_{\pi}^2}{\Lambda_{QCD}}\propto m_{quark}$. This $B_D$'s dependence is estimated considering a simple square well potential for deuteron. It is also estimated by uncertainty principle: $p\sim\frac{1}{\Delta x}\sim m_{\pi}$, and $-B_D+\frac{p^2}{2m_N}\sim 0$.

For the other neutron-induced reactions, $^3$He$(n,p)t$ and $^7$Be$(n,p)^7$Li, we use the rate fitted from the experimental data as found in Ref.~\cite{CFO} for the former and Ref.~\cite{SKM} for the latter, and they are assumed to be independent of the couplings because there are no theoretical derivation for these. Because $^3$He$(n,p)t$ affects $^3$He abundance which is not used for our analysis, our result does not change by neglecting its coupling dependence. On the other hand,  $^7$Be$(n,p)^7$Li affects $^7$Li abundance a lot when $\eta$ is large so its effect may be large. However, since there is large observational uncertainty in $^7$Li, it is expected that our result does not change but we have to know its coupling dependence to extract a reliable constraint from $^7$Li data when the uncertainty of $^7$Li abundances decreases in the future.

\section{The dilaton dependence of the coupling constants}\label{sec:dilaton}

Next, we introduce the action which governs the variation of the coupling constants as is used in Ref.~\cite{CampbellOlive}. This is the tree level low energy action of the heterotic string in the Einstein frame.
\begin{eqnarray}
S=\int d^4x\sqrt{-g}\biggl( \frac{1}{2\kappa^2}R-\frac{1}{2}\partial_{\mu}\Phi\partial_{\mu}\Phi-\frac{1}{2}\partial_{\mu}\phi\partial_{\mu}\phi \nonumber \\
{}-\Omega^{-2}V(\phi)
-\bar{\psi}\gamma^{\mu}D_{\mu}\psi-\Omega^{-1} m_{\psi}\bar{\psi}\psi \nonumber \\ {}-\frac{\alpha^{\prime}}{16\kappa^2}\Omega^2 F_{\mu\nu}F^{\mu\nu}\biggr). \label{eq:Eframeaction}
\end{eqnarray}
 where $\Phi$ is the dilaton field, $\phi$ is an arbitrary scalar field and $\psi$ is an arbitrary fermion. $D_{\mu}$ is the gauge covariant derivative corresponding to gauge fields with field strength $F_{\mu\nu}$. $\kappa^2=8\pi G$ and $\Omega=e^{-\kappa\Phi/\sqrt{2}}$ which is the conformal factor used to move from the string frame. Powers of $\Omega(\Phi)$ multiplying terms in the action indicate the dilaton dependence of the coupling constants and masses.

More concretely, $\phi$ is Higgs field and $V(\phi)$ is its potential which we assume to be given by hand (as is done in the Standard Model). The all over $\Omega$ factor before the scalar potential means that the Higgs vacuum expectation value $\langle H\rangle$ is independent of the dilaton so has same value for now and BBN time. $\langle H\rangle$ is taken to be constant in our calculation.

$F_{\mu\nu}$ is the gauge field with gauge group including $SU(3)\times SU(2)\times U(1)$. We define its Lagrangian density as $-\frac{1}{4 g^2}F_{\mu\nu}F^{\mu\nu}$ where $g$ is the unified coupling constant. Comparing with Eq.~(\ref{eq:Eframeaction}), 
\begin{equation}
\frac{1}{g^2(M_p^2)}=\frac{\alpha^{\prime}e^{-\sqrt{2}\kappa\Phi}}{4\kappa^2}\equiv\frac{\alpha^{\prime}S}{4\kappa},
\end{equation}
where we define
\begin{equation}
S\equiv\frac{e^{-\sqrt{2}\kappa\Phi}}{\kappa}.\label{eq:Sdef}
\end{equation}
 For each value of $S$, we can calculate the gauge coupling constants at the low energy using renormalization group equations. $\alpha_{EM}$ almost does not run so,
\begin{equation}
\alpha_{EM}(M_{BBN})\approx\alpha_{EM}(M_p)=\frac{g(M_p)^2}{4\pi}=\frac{\kappa}{\pi\alpha^{\prime}}S^{-1}.
\end{equation}
Therefore,
\begin{equation}
\frac{\alpha_{EM,bbn}}{\alpha_{EM,now}}=\left[\frac{S_{bbn}}{S_{now}}\right]^{-1}=\frac{1}{1+v_S}. \label{eq:rAlpha}
\end{equation}
where we define fractional $S$ variation,
\begin{equation}
v_S=\frac{S_{bbn}-S_{now}}{S_{now}}. \label{eq:varySdef}
\end{equation}
From the solution of 1-loop RGE for the SU(3) coupling constant $g_3$, of which integration constant  is determined by $g_3(\Lambda_{QCD})=\infty$,
\begin{equation}
\frac{\kappa}{\pi\alpha^{\prime}}S^{-1}=\frac{g_3(M_p)^2}{4\pi}\approx\frac{12\pi}{27\log(M_p^2/\Lambda_{QCD}^2)}\label{eq:rLQCD}
\end{equation}
or
\begin{equation}
\Lambda_{QCD}=M_p \exp\left(-\frac{2\pi^2\alpha^{\prime}S}{9\kappa}\right).
\end{equation}
Therefore, we obtain
\begin{eqnarray}
\frac{\Lambda_{QCD,bbn}}{\Lambda_{QCD,now}}&=& \exp\left(-\frac{2\pi^2\alpha^{\prime}[S_{bbn}-S_{now}]}{9\kappa}\right)\nonumber \\&=&\exp\left(-\frac{8\pi^2}{9g(M_p)^2}v_S\right),
\end{eqnarray}
where we use $g(M_p)^2=0.1$.

Finally, $\psi$'s are the ordinary SM leptons and quarks. As we take $\langle H\rangle=const.$, Yukawa couplings $y$ depend on dilaton as  $\propto e^{\kappa\Phi/\sqrt{2}}$. In terms of $S$,
\begin{equation}
\frac{y_{bbn}}{y_{now}}=\frac{1}{\sqrt{1+v_S}}.
\end{equation}

\section{Constraints on the variation of the coupling constants}\label{sec:constraint}

Using the model described in the previous section, we can express the coupling dependence of the BBN input parameters considered in Sec.~\ref{sec:coupling} with the fractional variation of the dilaton $v_S$ defined in Eqs.~(\ref{eq:Sdef}) and (\ref{eq:varySdef}). In order to quantify how much variation is consistent with the observation, we calculate the abundances for different values of $v_S$ in addition to $\eta$ with the standard BBN code \cite{Kawanocode}. Then we calculate $\chi^2(\eta,v_S)$ as,
\begin{equation}
\chi^2=\sum_i \frac{(a_i^{th}-a_i^{obs})^2}{(\sigma_i^{th})^2+(\sigma_i^{obs})^2}
\end{equation}
where $i$ is the type of the element with which we try to put a constraint. To estimate theoretical errors, we have performed 1000 Monte-Carlo simulations using the values of Ref.~\cite{CFO} for the nuclear reaction rates uncertainty and Particle Data Group \cite{PDG2000} for the neutron lifetime $885.7\pm0.8$ s. For the observational errors for D, $^4$He and $^7$Li, we adopt:
\begin{eqnarray}
({\rm D/H})^{obs}&=&(3.0\pm0.4)\times10^{-5} , \label{eq:Dadopted}\\
Y^{obs}&=&0.238\pm(0.002)_{stat}\pm(0.005)_{syst} \label{eq:He4adopted},\\
\log_{10}[(^7{\rm Li/H})^{obs}]&=&-9.76\pm(0.012)_{stat}\pm(0.05)_{syst}\nonumber  \\
& &{}\pm(0.3)_{add} \label{eq:Li7adopted}.
\end{eqnarray}
Eq.~(\ref{eq:Dadopted}) is taken from Ref.~\cite{O'MearaEtal2001} and Eq.~(\ref{eq:He4adopted}) from Ref.~\cite{FieldsOlive1998} where the first error is the statistical uncertainty and the second error is the systematic one. Eq.~(\ref{eq:Li7adopted}) is from  Ref.~\cite{BonifacioMolaro1997} with the error we have added for the uncertainty in chemical evolution \cite{FieldsEtal1996}.

\begin{figure}[t!]
\centering
\hspace*{-7mm}
\leavevmode\epsfysize=10cm \epsfbox{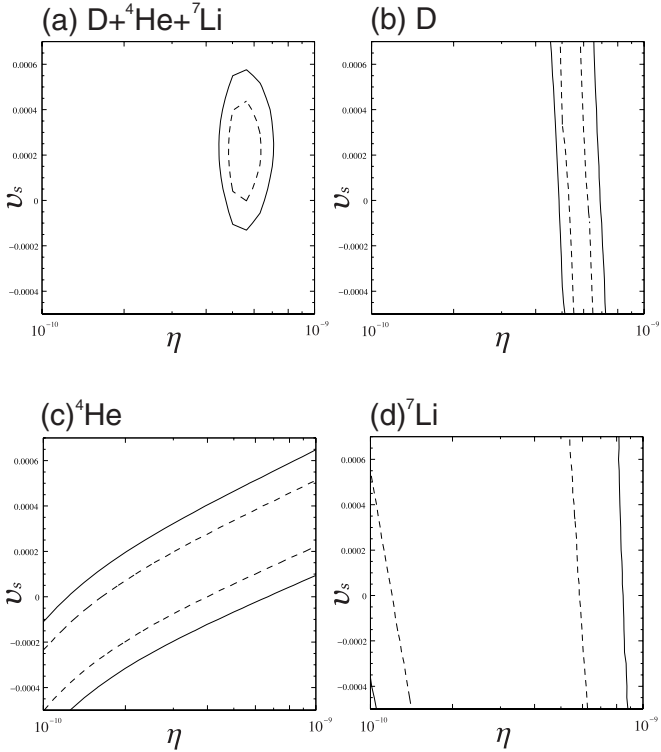}\\[2mm]
\caption{The contour lines on the $\eta-v_S$ plane drawn (a) using the data of the three light elements D, $^4$He and $^7$Li, (b) using D, (c) using $^4$He and (d) using $^7$Li. The vertical axis is $v_S=\frac{S_{bbn}-S_{now}}{S_{now}}$. The solid line is 95\% confidence level and the dotted line 70\%.}
\label{fig:CL}
\end{figure}

The results are shown in Fig.~\ref{fig:CL}. The shape of the contour lines are easily understood. The contours drawn from the $\chi^2$ of three elements (Fig.~\ref{fig:CL} (a)) are just the product set of D (Fig.~\ref{fig:CL} (b)) and $^4$He (Fig.~\ref{fig:CL} (c)) because $^7$Li (Fig.~\ref{fig:CL} (d)) does not give much constraint. 

D and $^7$Li do not vertically constrain much and $^4$He alone constrain $v_S$.  The reason is D and $^7$Li abundances are determined mainly by charged-particle reactions while $^4$He by freeze-out ratio. The former is determined by Coulomb barrier penetrability or electro-magnetic coupling $\alpha_{EM}$ and the latter by neutron-proton mass difference $\Delta m$. $\Delta m$ is depend linearly on $\Lambda_{QCD}$ as described in Sec.~\ref{sec:np-massdif}. $\alpha_{EM}$ and $\Lambda_{QCD}$ are related to each other through Eqs.~(\ref{eq:rAlpha}) and (\ref{eq:rLQCD}) which came from general notion of the gauge coupling unification. This relation tells that $\displaystyle \frac{\Delta \Lambda_{QCD}}{\Lambda_{QCD,now}}\approx\frac{80\pi^2}{9}\frac{\Delta\alpha_{EM}}{\alpha_{EM,now}}$ where $\Delta\alpha_{EM}=\alpha_{EM,bbn}-\alpha_{EM,now}$ etc.~so small increase in $\alpha_{EM}$ is accompanied by great increase in $\Lambda_{QCD}$ and $\Delta m$. This is the reason why $^4$He constrain $v_S$ much more than D and $^7$Li do in our model.

To make sure, we explain the trend of the $^4$He contour. As $v_S$ increases, $\Lambda_{QCD}$ decreases and hence $\Delta m$ increases. This makes $\frac{n_n}{n_p}|_{freezeout}$ decrease, so $^4$He abundance decreases. Since the $^4$He abundance is an increase function of $\eta$, the increase in $v_S$ relaxes the constraint for higher $\eta$. This shows up in the trend that the contour goes up in the direction of increasing $\eta$. 

The constraint on the dilaton field variation is obtained from Fig.~\ref{fig:CL} (a). For 95\% confidence level,
\begin{equation}
-1.5\times10^{-4} < v_S < 6.0\times10^{-4}, \label{eq:vsconstraint}
\end{equation}
or using $S=e^{-\sqrt{2}\kappa\Phi}/\kappa$ with $\kappa=(8\pi G)^{1/2}=(2.43\times10^{18}$GeV$)^{-1}$, we obtain
\begin{equation}
-1.0\times 10^{14}\ {\rm GeV} < \Delta\Phi < 2.6\times 10^{14}\ {\rm GeV}. \label{eq:dilatonconstraint}
\end{equation}

\begin{figure}[t!]
\centering
\hspace*{-7mm}
\leavevmode\epsfysize=5cm \epsfbox{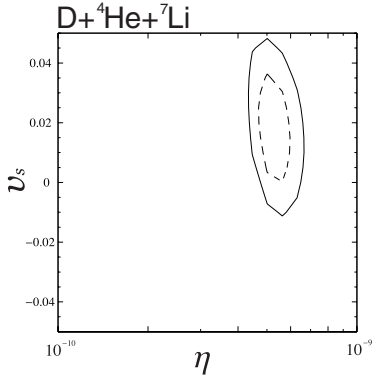}\\[2mm]
\caption{The contour lines when only $\alpha_{EM}$ is varied drawn using the data of the three light elements D, $^4$He and $^7$Li . The relation $v_S\approx-\frac{\Delta\alpha_{EM}}{\alpha_{EM}}$ holds.}
\label{fig:CLAlphaall}
\end{figure}
\section{Conclusion}

In summary, we have considered the BBN calculation with the coupling constants taking different values from the present ones and whose variation is determined by the dilaton field $\Phi$ as in the action (\ref{eq:Eframeaction}). The set up is the same as Ref.~\cite{CampbellOlive} but we have performed full nuclear reaction network calculation with the BBN code so the abundances of D and $^7$Li in addition to $^4$He are available to compare with the observation and make a constraint. This is done by calculating $\chi^2$ as a function of the baryon-to-photon ratio $\eta$ and the fractional variation of the dilaton $v_S$ defined in Eqs.~(\ref{eq:Sdef}) and (\ref{eq:varySdef}), and we have obtained the constraint (\ref{eq:vsconstraint}), which is the same order as the value found in Ref.~\cite{CampbellOlive}. It might be interesting that the standard BBN ($v_S$=0) is outside the allowed region with 70\% confidence level as seen in Fig.~\ref{fig:CL} (a) although it is not statistically significant.

The constraint on $v_S$ is readily converted to that on $\alpha_{EM}$ using Eq.(\ref{eq:rAlpha}), which becomes $\Delta\alpha_{EM}/\alpha_{EM}=-v_S$ when $v_S$ is small, and gives
\begin{equation}
-6.0\times10^{-4} < \frac{\Delta\alpha_{EM}}{\alpha_{EM}} < 1.5\times 10^{-4}, \label{eq:Alphaconstraint}
\end{equation}
which is two orders stricter than what found in Ref.~\cite{BIR} where only $\alpha_{EM}$ is varied. This feature should be common in the models in which variation of the gauge couplings are related to each other through gauge unification so the variation in $\Lambda_{QCD}$ is much larger than that of $\alpha_{EM}$.   

For comparison, we draw a contour diagram when only $\alpha_{EM}$ is varied. The result is
Fig.~\ref{fig:CLAlphaall}. The shape is similar but noticing the vertical scale, it is two orders larger than the Fig.~\ref{fig:CL} (a). The constraint becomes $-5.0\times10^{-2} < \Delta\alpha_{EM,only}/\alpha_{EM,only} < 1.0\times 10^{-2}$ and this is similar to the result obtained in Ref.~\cite{BIR} (factor difference is attributed to the difference in the adopted observational data). The contours in the Fig.~\ref{fig:CLAlphaall} completely covers the contours in the Fig.~\ref{fig:CL} (a). This is another demonstration that the constraint is not obtained by the variation of $\alpha_{EM}$ directly but by $\Lambda_{QCD}$.

Finally, we notice that the limit (\ref{eq:Alphaconstraint}) is consistent and has favourable trend with the result found in Ref.~\cite{QSO} from the quasar absorption systems, that is,  $\Delta\alpha_{EM,only}/\alpha_{EM,only}=(-0.72\pm0.18)\times 10^{-5}$. This shows $\alpha_{EM}$ was \textit{smaller} at $z=O(1)$. So if the variation is monotone, it should be further smaller at the BBN time. This is just the tendency of Eq.~(\ref{eq:Alphaconstraint}) (the allowed region is wider for the negative side). More quantitative comparison needs a model for time evolution of the coupling constants (for the model considered here, it is specifying dilaton potential) and the analysis of the quasar data under the situation where the couplings other than $\alpha_{EM}$ vary.

We have not reached a stage where such a constraint as (\ref{eq:dilatonconstraint}) can be used to say something about particle physics. In this paper, we bear string theory in mind and treat the time variation of the coupling constants as caused by the dilaton, scalar field found in the theory. However, in cosmology, there flourishes scalar fields to attack important problems (e.g.~inflation, quintessence) so the similar analysis (with different actions) can be used to constrain such scenarios. In order to find useful application, the analysis itself has to be sharpened because there still remains unsatisfactory estimates concerning the nuclear force. This sector will be improved with the field theoretic (QCD based) understanding of the nuclear force.

\end{document}